\documentclass[aps,prl,twocolumn,floatfix,preprintnumbers,nofootinbib,superscriptaddress,longbibliography,10pt]{revtex4-2}
\usepackage[margin=0.7in]{geometry}
\usepackage[utf8]{inputenc}

\usepackage{physics}
\usepackage{amsmath,amssymb,amsthm,amsfonts}
\usepackage[final]{graphicx}
\usepackage{subcaption}
\usepackage{lipsum}
\usepackage{mathtools}
\usepackage{enumitem,tabulary}
\usepackage[normalem]{ulem}
\usepackage{indentfirst}
\usepackage[english]{babel}
\usepackage{textcomp}
\usepackage{multirow}
\usepackage{tikz}
\usepackage{tipa}
\usepackage{CJKutf8}
\usepackage{feynmf}
\usepackage{slashed}
\usepackage{braket}
\usepackage[toc,page]{appendix}
\usepackage{url}
\usepackage{natbib}
\usepackage{graphicx}
\usepackage{mathrsfs}  
\usepackage{cancel}
\usepackage{array}
\usepackage{booktabs}
\usepackage{verbatim}
\usepackage{ragged2e}

\usetikzlibrary{positioning,decorations.pathmorphing,decorations.markings,arrows}
\usepackage{latexsym,amsfonts,color,amsthm}
\usepackage[compat=1.1.0]{tikz-feynman}
\RequirePackage[colorlinks=true
,urlcolor=purple
,anchorcolor=purple
,citecolor=purple
,filecolor=purple
,linkcolor=purple
,menucolor=purple
,linktocpage=true
,pdfproducer=medialab
,pdfa=true
]{hyperref}

\begin{document}

\begin{flushright}
\preprint{MI-HET-887}
\preprint{PITT-PACC-2611}
\preprint{CETUP2026-004}
\end{flushright}

\title{Muon Bremsstrahlung as a New Probe of Dark Sector at Neutrino Experiments}

\author{P. S. Bhupal Dev}
\email{bdev@wustl.edu}
\affiliation{Department of Physics and McDonnell Center for the Space Sciences, Washington University, St.~Louis, MO 63130, USA}
\affiliation{PRISMA$^{++}$ Cluster of Excellence \& Mainz Institute for Theoretical Physics, 
Johannes Gutenberg-Universit\"{a}t Mainz, 55099 Mainz, Germany}

\author{Bhaskar Dutta}
\email{dutta@tamu.edu}
\affiliation{Mitchell Institute for Fundamental Physics and Astronomy, Department of Physics and Astronomy, Texas A\&M University, College Station, TX 77843, USA}

\author{Aparajitha Karthikeyan}
\email{aparajitha\_96@tamu.edu}
\affiliation{Mitchell Institute for Fundamental Physics and Astronomy, Department of Physics and Astronomy, Texas A\&M University, College Station, TX 77843, USA}

\author{Mudit Rai}
\email{muditrai@umich.edu}
\affiliation{Mitchell Institute for Fundamental Physics and Astronomy, Department of Physics and Astronomy, Texas A\&M University, College Station, TX 77843, USA}
\affiliation{Leinweber Institute for Theoretical Physics, Department of Physics, University of Michigan, Ann Arbor, MI 48109, USA}

\author{Zahra Tabrizi}
\email{z\_tabrizi@pitt.edu}
\affiliation{PITT PACC, Department of Physics and Astronomy, University of Pittsburgh, Pittsburgh, PA 15260, USA}

\begin{abstract}

    We show that muon bremsstrahlung provides a new production mechanism for light new physics at accelerator neutrino experiments beyond the meson kinematic limit. The intense, highly collimated muon beam produced alongside the neutrino beam in meson decays provides a powerful source for the bremsstrahlung production of new physics when it impinges on the beam dump at these facilities. 
    As a benchmark scenario, we consider a muonphilic scalar coupled to Heavy Neutral Leptons (HNLs) and show that muon bremsstrahlung enables the production of HNLs with masses beyond the kinematic reach of meson decays. Using this new production mechanism and focusing on HNL mixing with tau neutrinos, we find that the upcoming neutrino experiments, such as the DUNE Near Detector, can probe previously unexplored parameter space for HNL masses up to $\mathcal{O}(1)~\mathrm{GeV}$ through their decay into pions, $\mu^+\mu^-$, and $e^+e^-$ final states at the detector. The resulting signals exhibit distinctive kinematics, allowing efficient discrimination from neutrino-induced Standard Model backgrounds. Our results demonstrate that the muon bremsstrahlung mechanism substantially extends the discovery potential of accelerator neutrino experiments for HNLs and other dark sector particles.

\end{abstract}

\maketitle

\section{Introduction}
\label{sec:Introduction}

There are compelling motivations for exploring physics beyond the Standard Model (BSM), including the unresolved nature of dark matter and the origin of neutrino masses. Many proposed BSM extensions present new particles that interact both among themselves and with the Standard Model (SM) particles. Over the past decades, a broad range of experimental and phenomenological probes have been developed to search for such scenarios, placing increasingly stringent constraints on new physics interactions. At the same time, ongoing and future experiments continue to provide novel opportunities to explore currently unconstrained regions of BSM parameter space across a wide range of masses and couplings.

Fixed-target accelerator neutrino facilities with high-intensity proton beams, such as MiniBooNE~\cite{MiniBooNE:2008paa}, MicroBooNE~\cite{MicroBooNE:2015bmn}, SBND~\cite{AlvarezGarrote:2024szs}, ICARUS~\cite{ICARUS:2004wqc}, and DUNE~\cite{DUNE:2016hlj,DUNE:2021tad}, offer a particularly promising environment for BSM searches. In addition to the intense hadronic and electromagnetic activity generated at the target, these facilities produce large fluxes of neutrinos, mesons, and charged leptons. The experimental setup typically consists of magnetic focusing horns located downstream of the target, which focus charged mesons such as pions and kaons. These mesons subsequently decay within a decay pipe, producing neutrinos and charged leptons before being absorbed in a massive beam dump or absorber. While the primary objective of the focusing system is to maximize the neutrino flux, the resulting meson and lepton beams also enable a rich program of BSM searches~\cite{Ilten:2022lfq,Batell:2022dpx,Gori:2022vri}.  Numerous studies have investigated exotic two-body and three-body meson decays in the decay region~\cite{Batell:2019nwo, Berryman:2019dme, Dutta:2021cip,MicroBooNE:2022ctm,Dev:2023rqb, Batell:2023mdn,Dutta:2023fnl}. Furthermore, signatures arising from neutrino scattering in the detector, beam dump, or intervening material have also been vastly investigated~\cite{Brdar:2020dpr, Dutta:2025npn, Dutta:2025sba}. Complementary production mechanisms, including proton bremsstrahlung~\cite{Foroughi-Abari:2021zbm, Foroughi-Abari:2024xlj, Kling:2025udr}, prompt neutral meson decays~\cite{Blumlein:2011mv}, Compton-like scattering, and Primakoff processes within electromagnetic showers~\cite{
Dent:2019ueq, Blinov:2024pza}, have significantly expanded the sensitivity of these facilities to a variety of new physics scenarios involving scalar and vector mediators, Heavy Neutral Leptons (HNLs), and dark matter. 

In this work, we propose muon bremsstrahlung from the secondary muon beam as a new production mechanism for BSM physics at neutrino beam-dump facilities, including MiniBooNE, SBND, ICARUS, and the future DUNE Near Detector (DUNE ND). Although muon bremsstrahlung is a well-established production mechanism in high energy muon interactions and forms the basis of dedicated muon-beam experiments such as NA64$\mu$~\cite{Tsai:1986tx,Tsai:1989vw,Rella:2022len,Chen:2018vkr,NA64:2024klw}, its potential at neutrino facilities has remained largely unexplored. We demonstrate that the intense, energetic secondary muon flux generated from focused meson decays provides a powerful probe of new physics beyond the meson decay limits through bremsstrahlung interactions in the beam absorber. While the accompanying neutrino beam continues toward the detector, the muons impinge on the absorber with energies of $\mathcal{O}(5-20~\mathrm{GeV})$ and a flux reaching $\sim0.1$ muons per proton-on-target in the LBNF beamline, making them an efficient source of new particles even within a single interaction length. An important feature of this new production mechanism is that muons originating from two-body meson decays typically carry a larger fraction of the parent meson energy than the accompanying neutrinos. This effect is particularly pronounced at neutrino facilities because the parent mesons are focused before decaying. In contrast, conventional beam-dump experiments such as SHiP~\cite{SHiP:2021nfo} and DarkQuest~\cite{Apyan:2022tsd} do not benefit from meson focusing, and the mesons are instead rapidly absorbed by a thick dump located immediately downstream of the target. This leads to substantially reduced energetic muon fluxes~\cite{Blinov:2024gcw,Blinov:2025aha},  compared to those available at neutrino facilities as studied here.

Muon bremsstrahlung production of new physics has previously been studied for muonphilic scalars at neutrino facilities~\cite{Cesarotti:2023udo} and at FASER$\nu$~\cite{MammenAbraham:2025gai}. However, searches based on visible scalar decays lose sensitivity for heavier scalars due to their shorter lifetimes, leaving the potential of muon bremsstrahlung to produce heavier new particles largely unexplored. To demonstrate this untapped potential, 
we consider a representative new physics scenario, namely, HNLs, as a benchmark. HNLs are well-motivated extensions of the SM that can explain neutrino masses through the seesaw mechanism~\cite{Minkowski:1977sc, Mohapatra:1979ia,Gell-Mann:1979vob,Yanagida:1979as}. Traditionally, HNL production at fixed-target experiments has been studied primarily through meson decays, where the HNL is produced via its mixing with active neutrinos~\cite{Asaka:2012bb,Kelly:2021xbv,MicroBooNE:2022ctm,MicroBooNE:2023eef, Alves:2024feq, Feng:2024zfe,Hostert:2025ffy}. 
More recently, additional production channels, including proton bremsstrahlung in gauge-extended models such as $B-L$~\cite{Ballett:2019pyw,Capozzi:2024pmh,Burk:2026mvk}, or through mixing with axion-like particles~\cite{Abdullahi:2023gdj} have also been explored. 
These frameworks are often motivated by ultraviolet-complete theories that contain HNLs together with additional scalar and vector mediators. A comprehensive exploration of such models therefore requires accounting for all relevant production channels across different mass and coupling regimes.

We show that bremsstrahlung emission of scalar or vector mediators from muons interacting within the absorber can generate a substantial HNL flux. The partitioning of the interaction strength between production and decay couplings allows the production rate and detectable lifetime to be varied separately to a useful extent, opening sensitivity to  
regions of parameter space that are inaccessible through conventional meson-decay searches, particularly for heavier HNL masses whose production via meson decay becomes kinematically suppressed. We further investigate the energy and angular distributions of the resulting HNL decay products at the detector, including channels such as $\nu e^+e^-$, $\nu\pi^0$, and $\nu \mu^+\mu^-$. These observables provide powerful discriminants for separating muon bremsstrahlung-induced HNL signals from SM neutrino-induced backgrounds.

\section{Models}
\label{sec:models}

In this study, as a benchmark model, we will demonstrate the potential of muon bremsstrahlung in searching for BSM models that contain a  muonphilic real scalar mediator ($\phi$) that also couples to HNLs ($N$). The scalar communicates to the SM and the dark sector through Yukawa interactions with the muons and HNLs, respectively. The relevant interaction Lagrangian is given as 
~\cite{Batell:2017kty, Dutta:2020scq,Harris:2022vnx}
\begin{equation}
      -\mathcal{L} \supset  y_{\mu}\bar{\mu}\mu \phi+ y_N \bar{N}N\phi\,.
\end{equation} 
Similar phenomenologies can be studied for vector and axial-vector mediators, where the anomaly-free model features couplings to other SM particles. Example models include $U(1)_{L_\mu-L_\tau}$~\cite{He:1991qd} or $U(1)_{T{3R}}$~\cite{Dutta:2019fxn} model with couplings to the right-handed muons and single-generation quarks. However, based on all the relevant SM couplings in the model, other production modes along with muon bremsstrahlung will contribute to the signal rates. Therefore, for simplicity we restrict this analysis to a scalar that couples only to the muons in the SM sector.

The HNL can mix with one or more SM flavor neutrinos in addition to its coupling to $\phi$, as they naturally arise from the seesaw solution to the neutrino masses~\cite{Minkowski:1977sc, Mohapatra:1979ia,Gell-Mann:1979vob,Yanagida:1979as}, or its variants like the inverse seesaw~\cite{Mohapatra:1986bd}. The presence of HNLs, therefore, modifies neutrino flavor eigenstates such that
\begin{equation}
    \nu_\alpha = \sum_{i=1}^3U_{\alpha i} \nu_i + U_{\alpha N} N\,,
\end{equation}
where $U_{\alpha i}$ are the PMNS mixing matrix elements connecting the mass and flavor eigenstates of light neutrinos, while $U_{\alpha N}$ is the mixing between active neutrinos and HNLs. Therefore, the HNL can interact via weak interactions, which are suppressed by the mixing $\propto |U_{\alpha N}|^2$. In this study, we assume that the HNL mixes only with the $\tau$ flavor, i.e., $U_{\tau N} \neq 0$ and $U_{e N} = U_{\mu N} = 0$, which is less constrained than the electron/muon mixing case. Such a single-flavor HNL mixing is a common assumption made in the phenomenological HNL
studies (see e.g., Refs.~\cite{Atre:2009rg,deGouvea:2015euy,Chrzaszcz:2019inj,Bolton:2019pcu,Fernandez-Martinez:2023phj}). The $\tau$-flavor
dominance assumed here is close to the $3\sigma$ allowed range of the normal-ordering scenario which allows up to
90\% $\tau$-component~\cite{Drewes:2022akb}.

\section{Phenomenology}
\label{sec:production}

\begin{figure*}[t]
    \includegraphics[width = 0.99\textwidth]{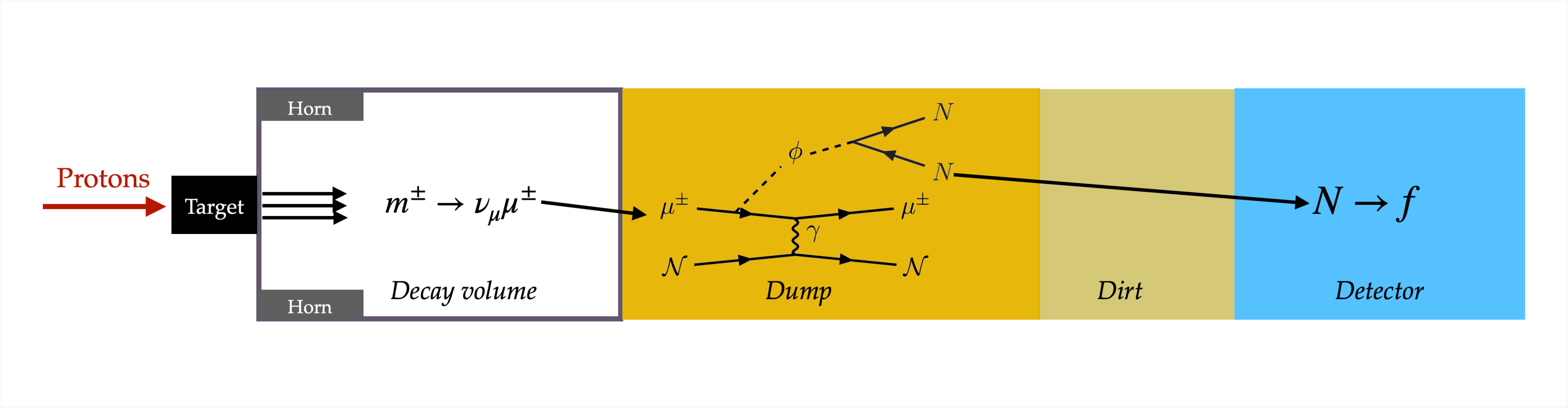}
    \captionsetup{justification=Justified, singlelinecheck=false}
    \caption{Schematic sketch (not to scale) showing the muon bremsstrahlung production of scalars ($\phi$) and, consequently, HNLs ($N$) at the dump  present at the end of the decay volume. 
    }
    \label{fig:sketch}
\end{figure*}
\noindent \textbf{\textit{Scalar Production:}} We now discuss the production of muonphilic scalars via muon bremsstrahlung at neutrino experiments. The setup is schematically shown in Fig.~\ref{fig:sketch}. High-energy protons impinge on a thin target, generating copious hadronic and electromagnetic activity. A significant fraction of long-lived charged mesons, particularly $\pi^{\pm}$ and $K^{\pm}$, survive the target and are focused along the beamline by magnetic horns. These mesons subsequently decay within the decay pipe, producing an intense focused neutrino beam, while the associated charged leptons, predominantly muons, eventually reach the absorber located at the end of the decay region. As these muons traverse $\mathcal{O}(10-100)~\mathrm{cm}$ in the absorber before losing a significant fraction of their energy, they can radiate muonphilic mediators through bremsstrahlung interactions. The emitted scalars subsequently decay into HNLs, giving rise to a novel production channel for HNLs at neutrino experiments.

We obtain the focused charged-meson flux at each experimental setup. For SBND, MiniBooNE, and ICARUS along the 8~GeV BNB setup, we use the {\tt RKHorn} simulation package~\cite{RKHorn2024}. For DUNE ND, we model the meson fluxes at the 120~GeV LBNF setup as a focused beam following Refs.~\cite{Dev:2023rqb,Dutta:2023fnl}. We then simulate the two-body decays of charged mesons in their rest frame and boost the decay products to the laboratory frame using the parent meson four-momenta distributions. For each meson, we account for the probability of decay within the decay pipe, whose length is approximately $50~\mathrm{m}$ and $220~\mathrm{m}$ for the BNB and LBNF beamlines, respectively. Assuming a muon is produced at $z_\mu$, we collect all muons that reach the absorber (assuming its cross-sectional area to be $4\times4~\mathrm{m}^2$) if the muon polar angle satisfies $\theta_{\rm dump}=\tan(r_{\rm dump}/(d_{\rm dump}-z_\mu))$. Since the dominant contribution arises from highly forward muons, we neglect the angular dependence of the bremsstrahlung cross section and integrate the muon flux over the absorber acceptance. Therefore, we use the muon flux $\frac{dN_\mu}{dE_\mu} = \int_0^{\theta_{\rm dump}} d\theta_\mu\ \frac{dN_\mu}{dE_\mu d\theta_\mu}$ and assume that they are all directed at $\theta_\mu = 0$. 

We use this muon flux to simulate scalar production through muon bremsstrahlung following the formalism of Ref.~\cite{Sieber:2023nkq}. The differential bremsstrahlung cross section is determined by the incident muon energy $E_\mu$, the scalar mass $m_\phi$, and the Yukawa coupling $y_\mu$. Since the bremsstrahlung rate increases with the energy of the incoming muon, scalar production is dominated by the energetic forward muons that reach the absorber from meson decays (see Fig.~\ref{fig:muonflux} in the {\it End Matter}). Folding the muon flux with the differential bremsstrahlung cross section yields the scalar flux:
\begin{equation}
    \begin{aligned}
        \frac{d^2N_{\phi}}{dE_{\phi}\,d\cos\theta_{\phi}} = \int &dE_{\mu}\, n_{\text{dump}}\,\lambda_{\mu}(E_{\mu})\frac{d^2\sigma_{\text{brem}}(E_{\mu})}{dx\,d\cos\theta_{\phi}} \\
        & \times J(x,E_{\phi})\frac{dN_{\mu}}{dE_{\mu}} \,,
    \end{aligned}
\end{equation}
where the differential cross section is expressed in terms of $x \equiv E_{\phi}/E_{\mu}$, the fraction of the muon energy transferred to the scalar, and the cosine of the angle between the scalar and the incoming muon, $\cos\theta_\phi$. The corresponding Jacobian is $J(x,E_{\phi}) = 1/E_{\mu}$. The number density of target nuclei in the dump is $n_{\rm dump}=N_{\rm A}\rho_{\rm dump}/A$, where $\rho_{\rm dump}$ is the dump density, $A$ is the atomic mass number of the target material, and $N_{\rm A}$ is the Avogadro number. We denote by $\lambda_\mu(E_\mu)$ the average distance traveled by a muon before it loses $10\%$ of its initial energy. This energy-dependent distance is obtained from the muon energy-loss equation~\cite{ParticleDataGroup:2024cfk}.

Muon-philic scalars can also be produced through the three-body decays of charged mesons, $K^\pm/\pi^\pm \rightarrow \mu^\pm + \nu_\mu + \phi.$
Unlike the corresponding two-body leptonic decays, these processes evade helicity suppression due to the available three-body phase space and can exhibit sizable branching ratios. Consequently, they constitute an important source of $\phi$ production for $m_\phi < m_{K^\pm/\pi^\pm}-m_\mu$. 

Since our focus is on the HNL mass and mixing parameter space, we fix $m_{\phi}/m_N=2.1$, and saturate existing bounds on $y_\mu$ from invisible-scalar searches at NA64$\mu$~\cite{NA64:2024klw}, BaBar4$\mu$~\cite{BaBar:2016sci}, and CMS4$\mu$~\cite{CMS:2024jyb}. Therefore, $y_{\mu}$ varies between $\sim [8\times10^{-4},2\times10^{-3}]$ for $10~\mathrm{MeV}\lesssim m_{\phi}\lesssim5~\mathrm{GeV}$. As these limits apply only to invisibly decaying scalars, we require $\mathrm{BR}(\phi\to NN)\simeq90\%$, typically corresponding to $y_N\gtrsim10\,y_{\mu}$.

\medskip

\noindent  \textbf{{\textit{HNL Production:}}} We next simulate scalar decays into HNLs. The fraction of scalars decaying into HNLs depends on the Yukawa couplings and the scalar boost factor. The scalar decay length in the lab frame is:
\begin{equation}
    \lambda_\phi = 1.74 \left( \frac{10^{-2}}{y_N} \right)^2 \left( \frac{1~\text{GeV}}{m_{\phi}}\right) \left( \frac{p_{\phi}}{m_{\phi}}\right)~\text{nm}\,.
\end{equation}
Therefore, we find that even for boost factors $p_{\phi}/m_{\phi}\sim 10^2-10^3$, the scalar decays within millimeters. We therefore treat the scalar as decaying promptly into HNLs at the dump, yielding 
\begin{equation}
    \begin{aligned}
        \frac{d^2N_{N}
        }{dE_{N}\,d\cos\theta_{N}} = \int &dE_{\phi}\,d\cos\theta_{\phi}\frac{d^2N_{\phi}(y_{\mu}, m_\phi) }{dE_{\phi}\,d\cos\theta_{\phi}} \\
        &\times \frac{d^2\text{BR}(y_{\mu}, y_N)}{dE_{N}\,d\cos\theta_{N}}\,,
    \end{aligned}
\end{equation}
where as mentioned above, the scalar mass is $m_\phi=2.1m_{N}$. 

\begin{figure}[t!]
    \raggedright
    \includegraphics[width = 0.48\textwidth]{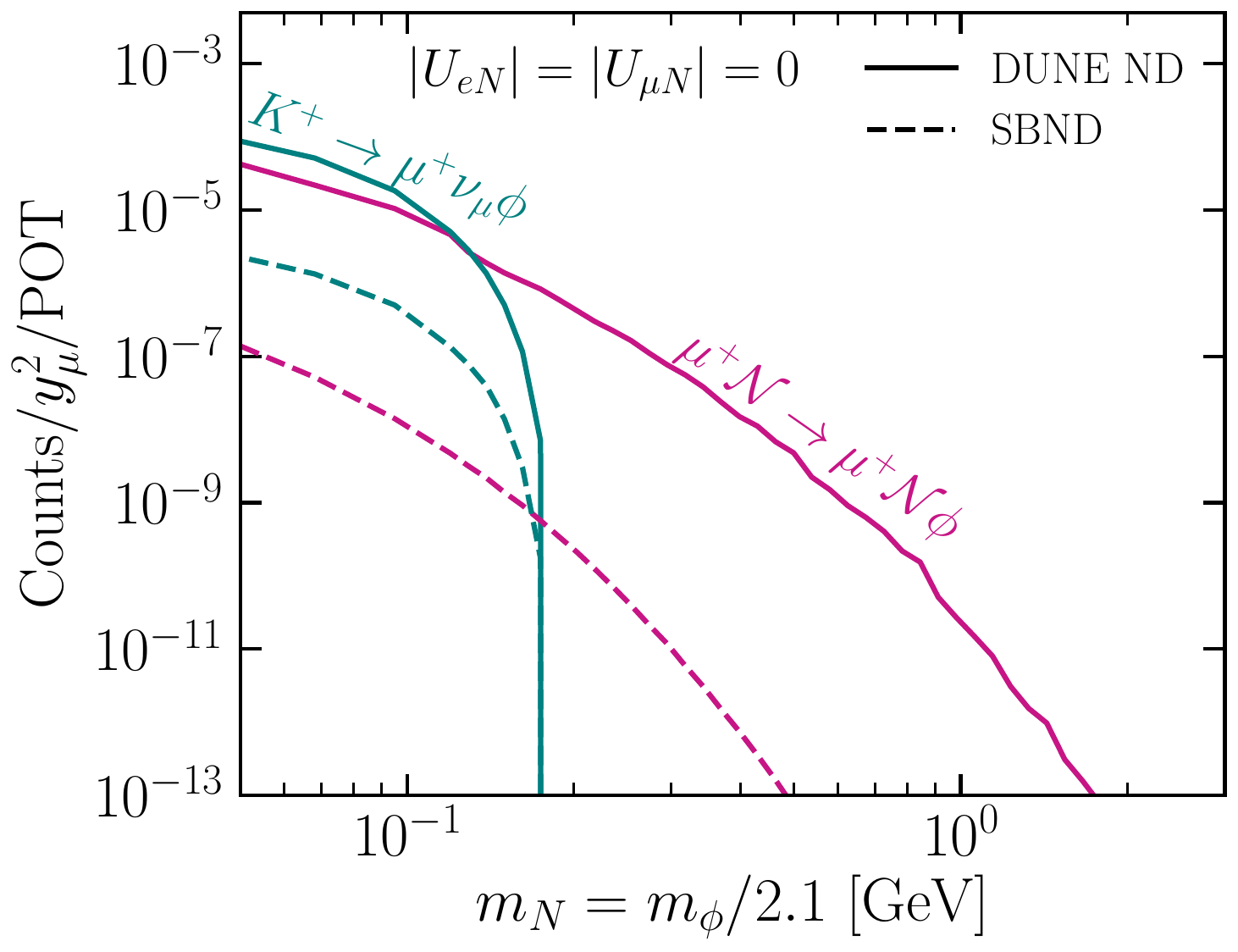}
    \captionsetup{justification=Justified, singlelinecheck=false}
    \caption{The number of HNLs that are directed towards DUNE ND and SBND per POT per $y_{\mu}^2$, assuming that the HNL mixes with the tau neutrino only. We note that the number of HNLs produced from muon bremsstrahlung is independent of $|U_{\tau N}|$. 
    }
    \label{fig:numberplot}
\end{figure}

Figure~\ref{fig:numberplot} depicts the number of HNLs produced in the dump and directed toward SBND and the DUNE ND. Since MiniBooNE and ICARUS operate at the same BNB facility as SBND, their HNL yields can be obtained by appropriately scaling the SBND fluxes. Under our assumption of $|U_{eN}|=|U_{\mu N}|=0$, HNL production through two-body charged-pion and kaon decays via mixing with $\nu_e$ and $\nu_\mu$ is absent. Consequently, HNLs are produced exclusively through the muonphilic scalar, yielding a flux proportional to $y_\mu^2$. For $m_N>50~\mathrm{MeV}$ ($m_\phi>105~\mathrm{MeV}$), we compare production from charged-kaon three-body decays and muon bremsstrahlung. The former is kinematically limited to $m_\phi<m_{K^+}-m_{\mu^+}$, whereas bremsstrahlung accesses significantly heavier scalars since it is determined by the muon energy rather than a fixed hadronic mass scale. Unlike meson decays, the ratio of HNL yields at DUNE ND and SBND is mass dependent due to the more energetic muons in the 120 GeV LBNF beam, which enhance production of heavier HNLs. If $U_{eN}\neq0$ or $U_{\mu N}\neq0$, additional low-mass HNLs are produced through two-body charged-meson decays, such as $K^+/\pi^+\to\nu_{e/\mu}+e^+/\mu^+$, although these channels are limited to $m_N\lesssim492~\mathrm{MeV}$. While HNLs from $D^\pm\to e^\pm/\mu^\pm/\tau^\pm+N$ can kinematically reach GeV masses, the $D$-meson flux at LBNF is highly suppressed, making this contribution subdominant to muon bremsstrahlung.

Although we have considered a muonphilic scalar, we find similar or larger fractions for muonphilic vector and axial-vector gauge bosons that reach the detector. This is because the muon bremsstrahlung cross-section is around one-order of magnitude larger for vectors and axial vectors, accounting for multiple spin possibilities.

\medskip

\noindent  \textbf{{\textit{HNL Detection:}}} 
After obtaining the HNL flux at the dump, we compute the probability that the HNL reaches the detector and then decays into final states within the detector volume, which is given by: 
\begin{equation}
    \begin{aligned}
        P_{\text{decay}}(p_N, m_N) &= e^{-(d_\text{det} - d_\text{dump})/\lambda_N}(1 - e^{-\Delta_\text{det} /\lambda_N})\,, 
    \end{aligned}
\end{equation}
where $d_\text{det}$ and $d_\text{dump}$ are the positions of the detector and dump w.r.t the target, respectively, and $\Delta_\text{det}$ is the depth of the detector. Here $\lambda_N$ is the HNL decay length, given by
\begin{equation}
    \lambda_N \simeq 1~\text{m}  \left( \frac{0.2\times 10^{-15}~\text{GeV}}{\Gamma_{N}^{\text{weak}} + \Gamma_{N}^{\phi}} \right) \times \frac{p_N}{m_N}\,.
\end{equation}
Here,  $\Gamma_{N}^{\text{weak}}$
is the HNL decay width that receives contributions from the SM weak interactions via mixing, scaling as $|U_{\alpha N}|^2 G_F^2$, for which we use existing results~\cite{Bondarenko:2018ptm, Ballett:2019bgd, Coloma:2020lgy, Capozzi:2024pmh}. The HNL also decays via the muon-philic mediator, with decay width $\Gamma_{N}^{\phi}$, which scales as $|U_{\alpha N}|^2 (y_{\mu}y_N/m_{\phi}^2)^2$.

Finally, the number of signal events at the detector for a given final state $f$ is given as
\begin{equation}
    \begin{aligned}
        N_f =& \int dE_N\,d\cos\theta_N \frac{d^2N_{N} (y_{\mu}, m_{N})}{dE_{N}\,d\cos\theta_{N}} \Theta(\theta_N-\theta_{\text{det}}) \\ &
        \times P_{\text{decay}}(\sqrt{E_N^2 - m_N^2}, m_N)~\text{BR}(N\to f)\,,
    \end{aligned}
\end{equation}
where $\text{BR}(N\to f)$ is the branching ratio of the HNL to a given final state. In this study, we investigate the HNL decays to $e^+ e^-$, $\pi^0$, and $\mu^+ \mu^-$. The relevant HNL branching ratios before and after including the muonphilic scalar are given in Fig.~\ref{fig:brrelevant} in the {\it End Matter}. We perform a full Monte Carlo simulation of the signal chain, from convolution of the differential cross section with the muon flux through HNL production and two-body decays, their propagation and survival to the detector, and their decays within the detector volume.


\begin{figure*}[t]
    \RaggedRight 
    \includegraphics[width = \textwidth]{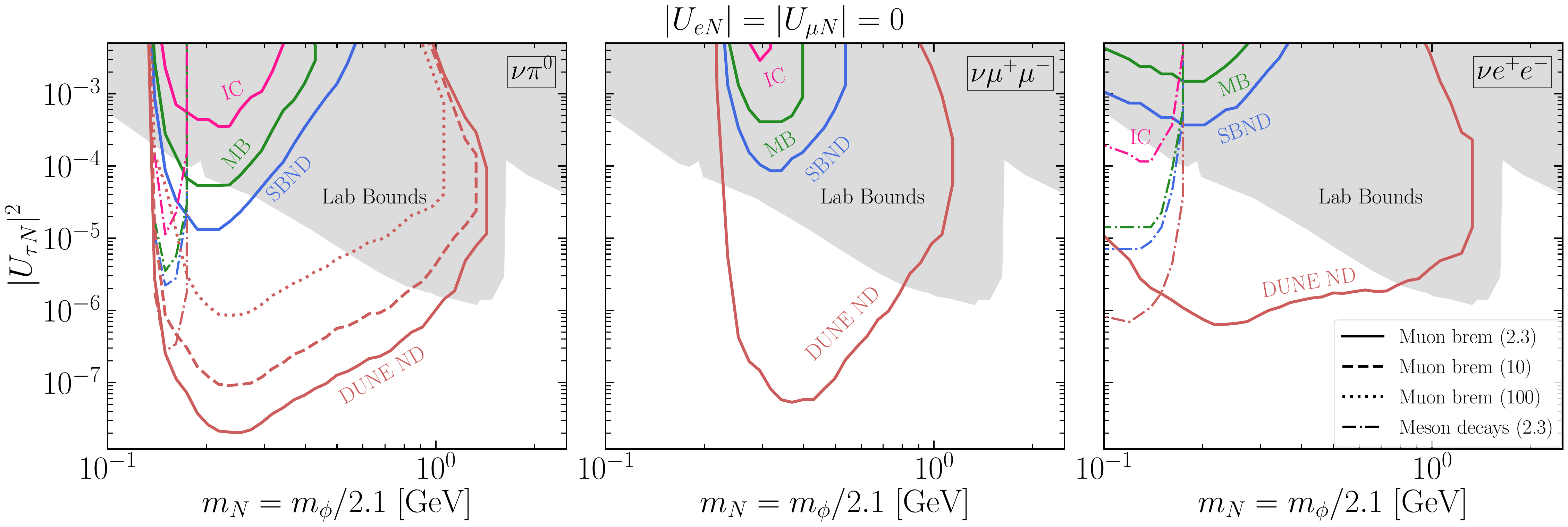}
    \captionsetup{justification=Justified, singlelinecheck=false}
    \caption{HNL sensitivity at DUNE ND, SBND, MiniBooNE (MB), and ICARUS (IC) in the $\nu \pi^0$ (left), the $\nu \mu^+ \mu^-$ (middle) and $e^+e^-$ (right) decay channels. The gray-shaded regions correspond to the existing constraints. See text for details.} 
    \label{fig:sens}
\end{figure*}


\section{Sensitivities}
\label{sec:Sensitivities}

We summarize our main results in Fig.~\ref{fig:sens}, which shows the HNL sensitivities at DUNE ND, SBND, MiniBooNE, and ICARUS to the single $\pi^0$ (left panel), $\mu^+ \mu^-$ (middle panel), and $e^+ e^-$ (right panel) topologies. The sensitivity to HNLs produced from muon bremsstrahlung is shown by the solid, dashed, and dotted lines for DUNE ND, corresponding to 2.3, 10 and 100 signal events, respectively, whereas for ICARUS, MiniBooNE and ICARUS, only the 2.3 event contours are shown. See {\it End Matter} and Fig.~\ref{fig:pi0bkg} for more details on the choice of these benchmark event numbers. Furthermore, for the single $\pi^0$ and $e^+ e^-$ signals, we also show by the thin dashed-dotted lines the sensitivity of HNLs produced from muonphilic scalars that arise from the meson decays, namely the three-body kaon decay for $m_{\phi} (2.1 m_N) < m_{K^+} - m_{\mu^+} (370~\rm{MeV})$. We note that the charged pion three-body decay does not contribute to the mass regime where the HNL can kinematically decay to $\nu \pi^0$. Furthermore, the HNLs from scalars produced from meson three-body decays are not heavy enough to lead to $\nu_\tau \mu^+ \mu^-$ decays. The gray shaded region represents existing constraints in this mass range from CHARM~\cite{Orloff:2002de, Boiarska:2021yho}, BEBC~\cite{WA66:1985mfx, Barouki:2022bkt}, BaBar~\cite{BaBar:2022cqj}, ArgoNeuT~\cite{ArgoNeuT:2021clc}, and DELPHI~\cite{DELPHI:1996qcc}, as summarized in Refs.~\cite{boltonHNL, hostertHNL}.  There exist additional constraints on the HNL parameter space from  BBN~\cite{Gelmini:2020ekg, Sabti:2020yrt, Boyarsky:2020dzc, Dev:2025pru} and SN 1987A~\cite{Mastrototaro:2019vug, Carenza:2023old}; however, the presence of the muonphilic scalar necessitates a dedicated study (similar to Refs.~\cite{Croon:2020lrf, Caputo:2021rux}) to reevaluate these constraints.  Furthermore, these constraints carry inherent uncertainties~\cite{Brdar:2026xlm}: BBN constraints are sensitive to cosmological models and reheating temperature, while those from SN 1987A depend on the progenitor model and the supernova envelope size. For these reasons, we do not show  these constraints in Fig.~\ref{fig:sens}. In fact, the laboratory-based probes such as the muon bremsstrahlung signature discussed here would provide an independent test of various cosmological and astrophysical models.

We find that muon bremsstrahlung can extend HNL searches to much larger mass regions. For DUNE ND, it can extend the sensitivities up to $m_N \simeq 1~\rm{GeV}$ and probe unexplored parameter space in all three channels shown in Fig.~\ref{fig:sens}. The enhanced HNL branching ratio into $\mu^+ \mu^-$ from the muonphilic scalar is particular to this model and can lead to clean signatures in a relatively background-free environment.  For the single-$\pi^0$ final state (left panel), both SBND and MiniBooNE can start to probe new parameter space as well. While MiniBooNE has observed a low-energy excess in the neutrino and anti-neutrino modes~\cite{MiniBooNE:2008yuf,MiniBooNE:2018esg,MiniBooNE:2020pnu}, muon bremsstrahlung related models are not helpful in explaining them because the resulting signal is much higher in energy than that required for explaining the MiniBooNE anomaly. 

The SBND results obtained here can be scaled to obtain the corresponding MicroBooNE sensitivities by accounting for the larger baseline, smaller detector volume, and lower accumulated POT, resulting in an overall suppression factor of around two-orders of magnitude in the expected number of events.


\section{Discussion and Conclusions}
\label{sec:Conclusions}


In this study, we show how muon bremsstrahlung can be a powerful probe of new physics sectors, particularly at the DUNE ND. Beyond HNLs, the muon-bremsstrahlung production mechanism can be readily extended to a broader class of BSM searches such as elastic/inelastic dark matter scenarios. More generally, studying new physics production through muon bremsstrahlung at neutrino facilities establishes a conceptual connection to future muon collider experiments~\cite{Accettura:2023ked}, where intense, high-energy muon beams may offer unprecedented sensitivity to HNLs and a wide range of hidden-sector particles.

In the example of this work, we have considered models with muon couplings only. However, if the model under consideration also contains a light mediator coupled to first-generation quarks, proton bremsstrahlung can provide an additional contribution to the HNL flux at higher masses, primarily through vector-meson resonance production. The signals induced by muon bremsstrahlung and proton bremsstrahlung can, in principle, be distinguished through their kinematic characteristics. Proton bremsstrahlung typically produces higher-energy, strongly forward-directed particles, resulting in a more collimated signal. In contrast, muon bremsstrahlung generally yields a softer and less forward-peaked spectrum. Consequently, muon bremsstrahlung serves as a complementary production channel that can enhance the sensitivity of neutrino experiments to new physics and potentially probe regions of parameter space inaccessible through proton-induced processes alone. 


\section*{Acknowledgements}

We would like to thank Francis Burk, Matheus Hostert, Gordan Krnjaic, Pedro Machado, Francisco Javier Nicolas-Arnaldos, Vishvas Pandey, and Mary Hall Reno for helpful discussions. The work of P.S.B.D. was supported in part by the US Department of Energy under grant No. DE-SC0017987 and by a Humboldt Fellowship from the Alexander von Humboldt Foundation. The work of B.D., A.K., and M.R. is supported by the U.S.~Department of Energy grant no. DE-SC0010813. The work of ZT is supported by Pitt PACC. For facilitating portions of this research, the authors wish to acknowledge the Center for Theoretical Underground Physics and Related Areas (CETUP*), The Institute for Underground Science at Sanford Underground Research Facility (SURF), and the South Dakota Science and Technology Authority for hospitality and financial support, as well as for providing a stimulating environment.

\bibliography{references}

\appendix
\section{END MATTER}

\section{A. Muon flux}
\begin{figure}[h]
    \raggedright
    \includegraphics[width=0.47\textwidth]{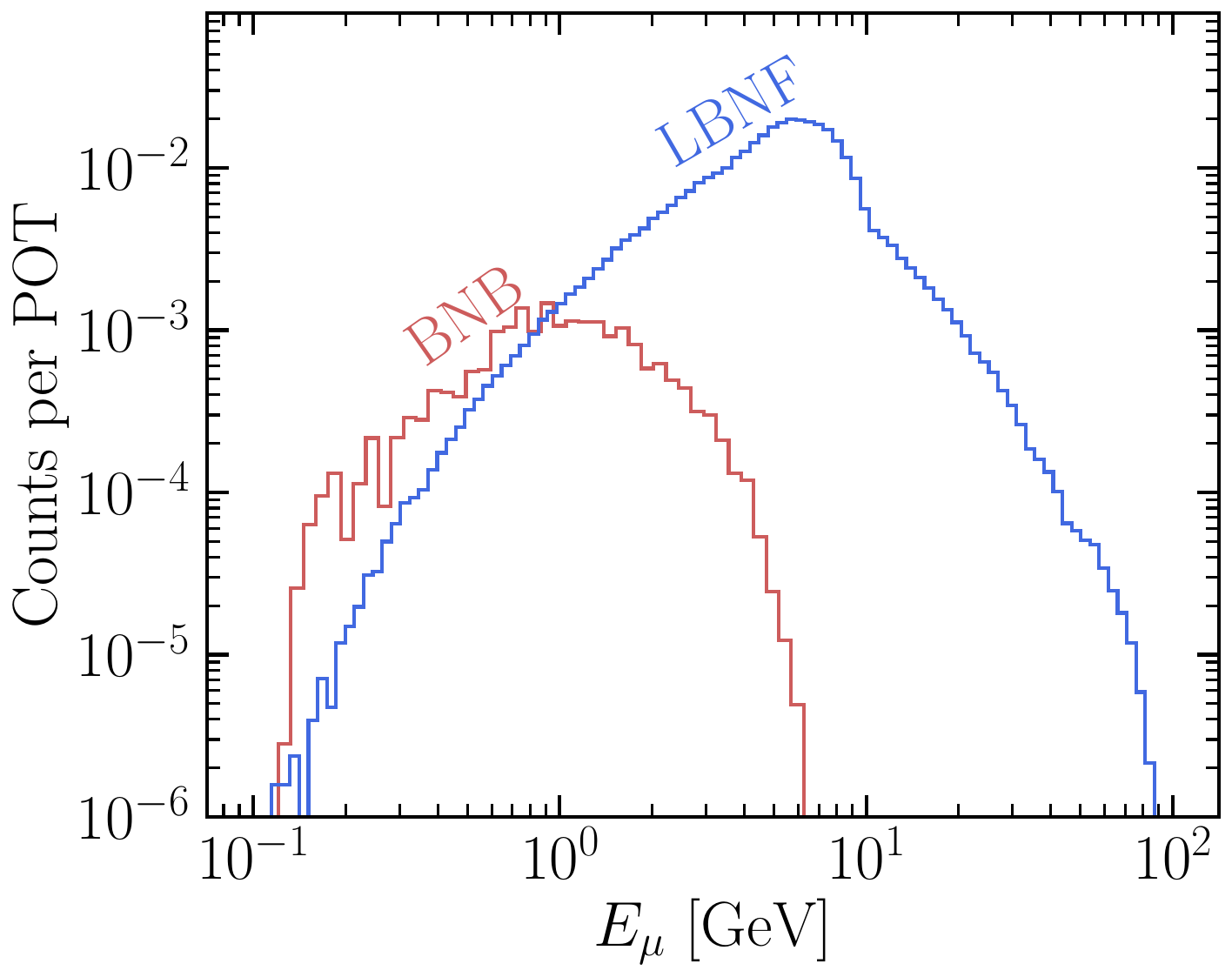}
    \captionsetup{justification=Justified, singlelinecheck=false}
    \caption{Muon flux that reaches the dump at the BNB and LBNF setups.}
    \label{fig:muonflux}
\end{figure}
Figure~\ref{fig:muonflux} depicts the energy distribution of the muons that reach the dump/absorber at the BNB and LBNF beamlines. Here, we see that the peak energy of muons at the dump are much higher than those of neutrinos. For BNB, the peak energy of muons is around 1~GeV, whereas the neutrino energies peak at around 700~MeV. Similarly, at LBNF, the muons have a peak energy of $\sim 9~\rm{GeV}$, whereas neutrinos peak at $\sim 5~\rm{GeV}$. This is an artifact of the two body decays of mesons, where the muons are much heavier than neutrinos. Furthermore, we find that there are a considerable amount of muons per POT that can contribute to muon bremsstrahlung production of new physics scenarios.


\section{B. HNL branching ratios}
\begin{figure}[h]
    \raggedright
    \includegraphics[width = 0.48\textwidth]{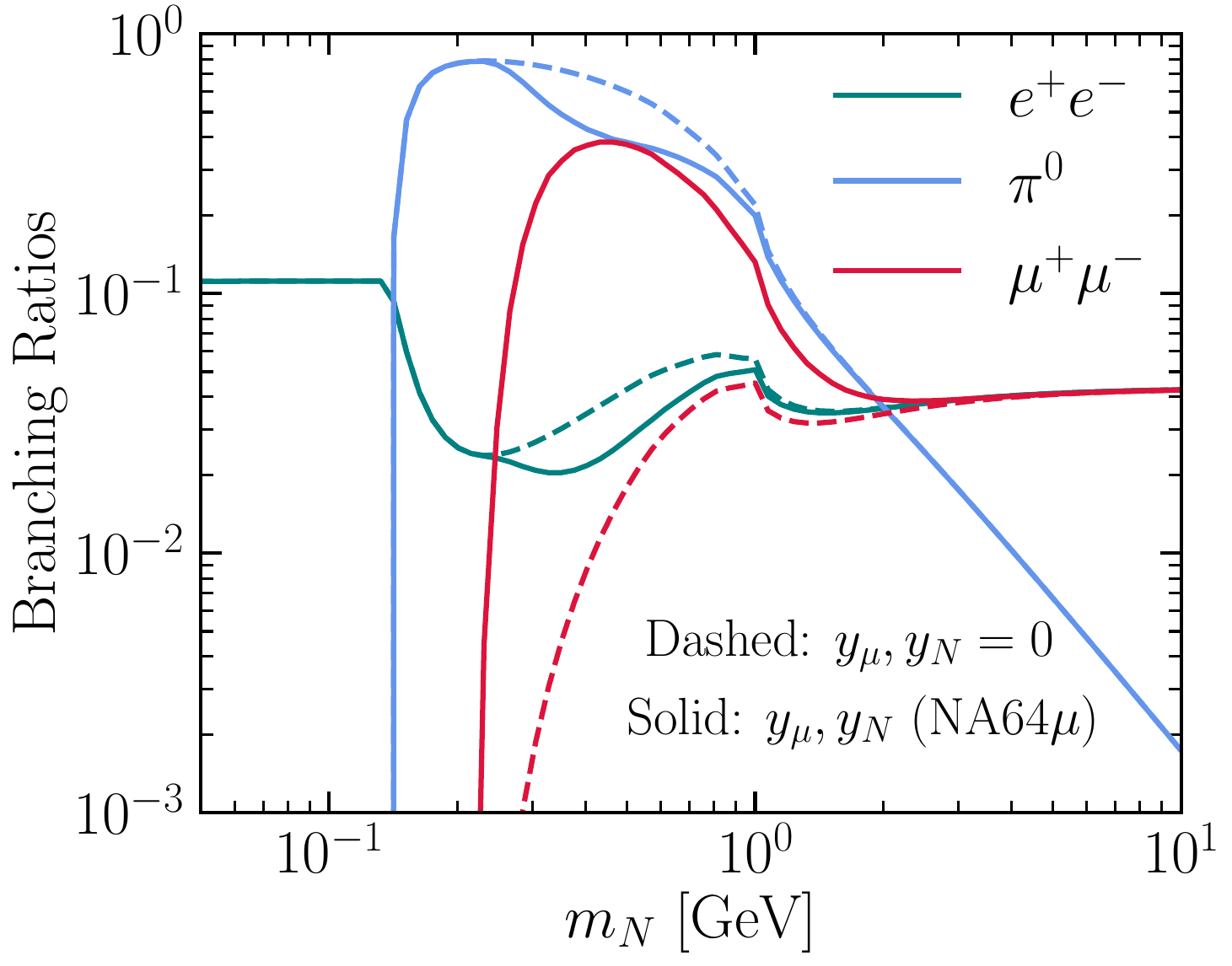}
    \captionsetup{justification=Justified, singlelinecheck=false}
    \caption{Relevant branching ratios of the HNL to $e^+e^-$, $\pi^0$ and $\mu^+\mu^-$. } 
    \label{fig:brrelevant}
\end{figure}
Figure~\ref{fig:brrelevant} shows the relevant branching ratios of HNLs to $\nu e^+ e^-$, $\nu \pi^0$, and $\nu \mu^+ \mu^-$ as a function of HNL mass. We show the branching ratios for cases when HNL decays only through mixing with neutrinos (through weak mediators) in dashed lines, as well as those with the additional presence of the muonphilic scalar mediator in solid lines. Due to the muonphilic mediator, we see an enhancement in the branching ratio to muon-antimuons as compared to the mixing-only scenario. Furthermore, we see a slight suppression in its branching ratio to electron-positron and neutral pions. This results in an enhanced sensitivity to muon-antimuons as seen in Fig.~\ref{fig:sens}.

\section{C. Signal vs. Background}
\begin{figure}[t]
    \includegraphics[width=0.49\textwidth, clip]{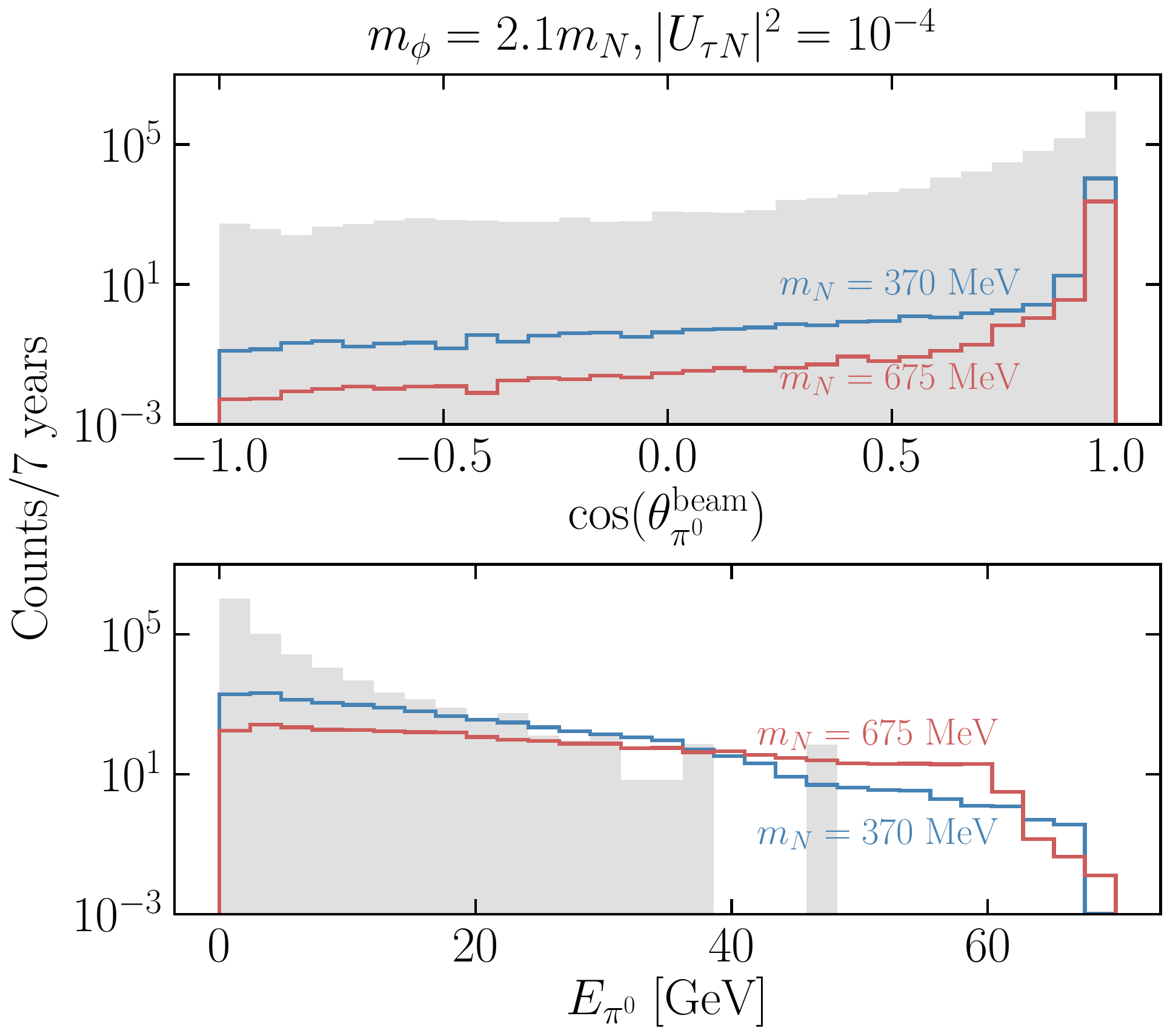} 
     \captionsetup{justification=Justified, singlelinecheck=false}
    \caption{Single $\pi^0$ event versus background distributions for two benchmark HNL masses $m_N = 370, 675~\rm{MeV}$ and mixing $|U_{\tau N}|^2 = 10^{-4}$. We also assume that the HNLs are produced from a scalar (with mass $m_\phi = 2.1 m_N$) that is produced via muon bremsstrahlung. }
     \label{fig:pi0bkg}
\end{figure}

To show how the SM backgrounds can affect our sensitivity analysis, for the DUNE ND we show in Fig.~\ref{fig:sens} the sensitivities for 2.3 (which corresponds to the background-free case), 10, and 100 event benchmarks. Focusing on the HNL decay into a single $\pi^0$ (left panel of Fig.~\ref{fig:sens}), the main source of the SM background would be the neutral current $\pi^0$ production from neutrino interactions at the detector. 
However, we find that the signal induced by the HNL decay is much more forward and energetic in comparison to the SM background. This is shown in Fig.~\ref{fig:pi0bkg}, where the upper panel corresponds to the cosine of the angle of the $\pi^0$ w.r.t the beamline, and the lower panel shows its energy distribution. The shaded histograms show the distributions for the background events, and the solid lines correspond to those of the HNL signal. Therefore, we see that by making appropriate upper bounds on $\theta_{\pi^0}^{\text{beam}}$ and lower bounds on $E_{\pi^0}$, the  backgrounds can be reduced from $\mathcal{O}(10^6)$ to $\mathcal{O}(10^3)$, while simultaneously retaining $10-50\%$ signal efficiency. As a result, assuming only statistical uncertainties, we estimate that approximately 100 signal events are required to achieve a $\sim2\sigma$ sensitivity to the HNL. The sensitivities could potentially be further extended through the use of machine-learning techniques to maintain larger signal efficiencies and suppress backgrounds further. However, a dedicated study in close collaboration with experimentalists is necessary to quantify the improvement. 

We show the sensitivities for the HNL decay to  $\nu \mu^+ \mu^-$ in the center panel of Fig.~\ref{fig:sens}. Here, we emphasize that the inclusion of the muonphilic scalar mediator significantly enhances the decay of HNL into $\nu \mu^+ \mu^-$ as compared to the weak-interaction-only case (see Fig.~\ref{fig:brrelevant}).  
Here, the main source of the SM background is the muon-neutrino charged current scattering, which can be reduced by appropriate cuts on the transverse momentum $p_T$ and the opening angle of the $\mu^+\mu^-$ pair. Therefore, we have used the kinematic cuts from Ref.~\cite{Coloma:2023oxx}, according to which we can practically get rid of all the $\mu^+\mu^-$ background by imposing $p_T^{\rm{tot}} < 125~\rm{MeV}$, and $1^{0} < \Delta \theta_{\mu\mu} < 40^{0}$. Since the HNLs are very energetic, the $\mu^+ \mu^-$ signal is quite forward with small opening angles, and small $p_T$ values. Therefore, we find that the signal efficiency can be retained to a great extent. We apply these cuts to the signal and plot the sensitivity for 2.3 events in Fig.~\ref{fig:sens}, corresponding to 90\% C.L. for background-free estimates.

Finally, for the case of HNL decay into $e^+ e^-$ final states, shown in the right panel of Fig.~\ref{fig:sens}, as Ref.~\cite{Coloma:2023oxx} shows, for the DUNE ND after imposing $\theta_{e^+e^-}^{\rm{beam}} < 5^0$, and $1^{0} < \Delta \theta_{ee} < 20^{0}$ cuts there will be approximately 900 background events left for 7 years of exposure. However, we note that these backgrounds are much softer compared to the signal coming from the HNL decay. Hence, by further requiring $E_{e^+ e^-} \gtrsim 3~\rm{GeV}$, the background can be reduced to zero. Correspondingly, we show the sensitivity contours for 2.3 events at the DUNE ND.

The sensitivities for SBND, MiniBooNE, and ICARUS are all shown by the 2.3 event contours. While we do not perform a detailed signal versus background analysis for these detectors, the HNLs produced from muon bremsstrahlung result in high energy ($E^N_{\text{vis}} \gtrsim 1-2~\text{GeV}$) and forward signals at the detectors. Therefore, there is a stark difference between these signals and the neutrino-induced backgrounds which are much lower in energy ($E^{\nu}_{\text{vis}} \simeq 0.5-1~\text{GeV}$)~\cite{Pandey}. We predict that a detailed analysis would result in a sensitivity line that is very close to the background-free predictions shown in Fig.~\ref{fig:sens}. 

\end{document}